\documentclass[11pt]{article}
\usepackage{vietnam,epsfig}

\newcommand{\dzero}     {D\O\  }

\newcommand{\met}       {\mbox{$\not\!\!E_T$}}

\def\be{\begin{equation}}
\def\ee{\end{equation}}
\def\bea{\begin{eqnarray}}
\def\eea{\end{eqnarray}}

\begin{document}
\vspace*{4cm}
\title{Search for Single Top Production at the Tevatron}

\author{ Reinhard Schwienhorst\footnote{On behalf of the \dzero and CDF collaborations.} }

\address{Department of Physics and Astronomy, Michigan State University, East Lansing, MI 48824. \\
	Email: schwier@fnal.gov}

\maketitle\abstracts{
Searches for the electroweak production of single top quarks
have been started at the Fermilab Tevatron proton-antiproton collider
using Run~II data by both the \dzero and CDF collaborations.  
Using a dataset of approximately
$160{\rm pb^{-1}}$, neither experiment finds evidence for Single Top production
and sets 95\% C.L. upper limits on the production cross section. The \dzero 
limits are 19pb on the s-channel production, 25pb on the t-channel production,
and 23pb on the combined s+t-channel production. The CDF limits are
8.5pb on the t-channel production and 13.7pb on the combined s+t-channel production. }

\section{Introduction}
The top quark, discovered by the Tevatron CDF and {\dzero}
collaborations~\cite{Abe:1995hr,Abachi:1995iq} in top pair production
is the heaviest elementary particle found so far. Though top quarks are predominantly produced
in pairs via the strong interaction at the Tevatron, they can also be produced singly
in electroweak interactions (see Ref.~\cite{Heinson:1996zm} and references therein). 
This production mechanism allows for a direct measurement of the CKM mixing angle 
$|V_{tb}|$. It also is sensitive to Physics beyond the Standard Model\cite{Tait:2000sh}.

\begin{figure}
\begin{minipage}{0.5\textwidth}
  \centering
  \psfig{figure=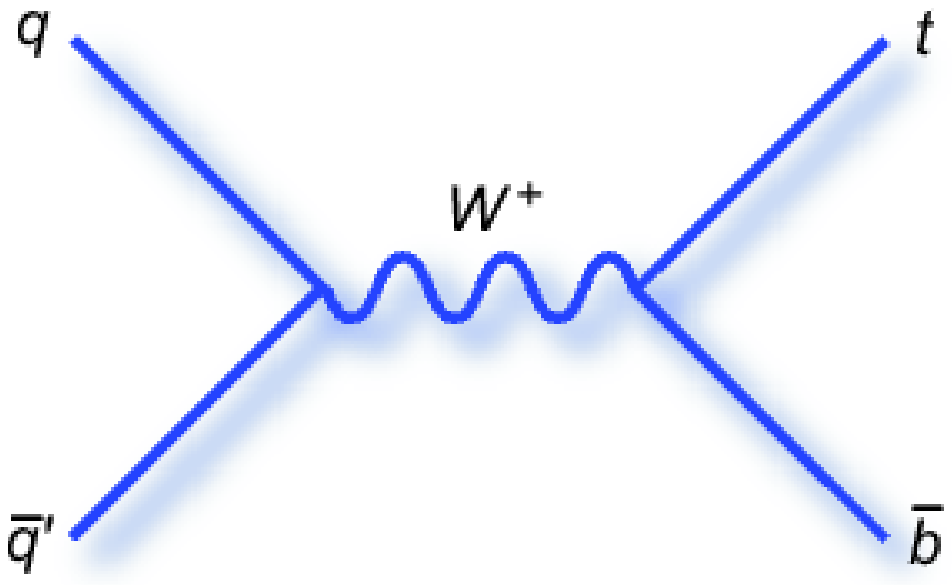,height=1.5in}
\end{minipage}
\begin{minipage}{0.5\textwidth}
  \centering
  \psfig{figure=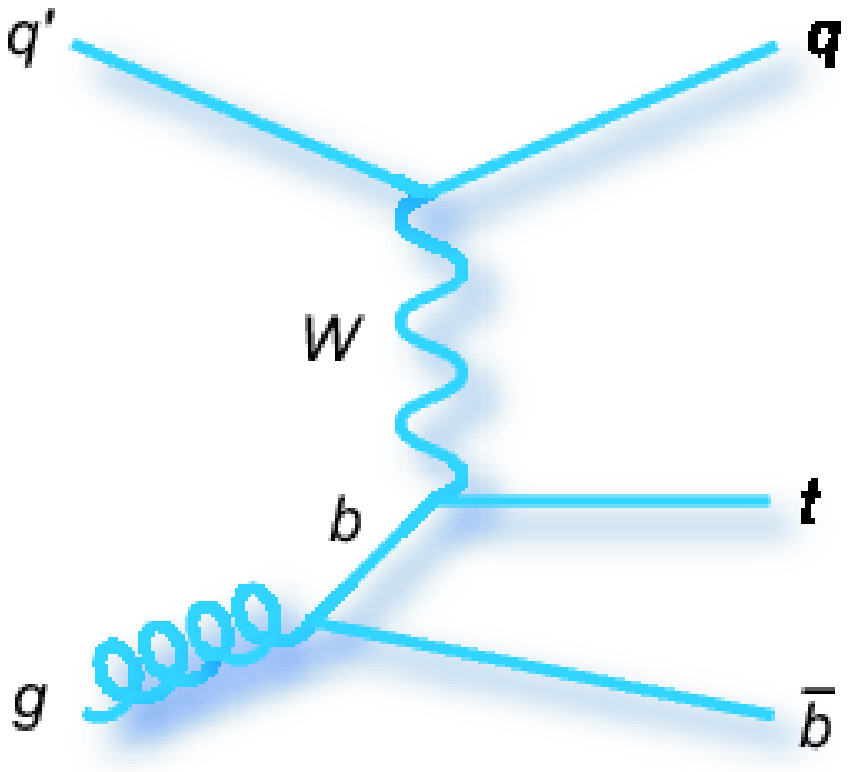,height=1.5in}
\end{minipage}
\caption{The dominant Feynman diagrams for $s$-channel
(left) and $t$-channel (right) single top quark production at the Tevatron.
\label{fig:feynmantbtqb}}
\end{figure}

The two main production modes at the Tevatron are the s-channel and t-channel exchange
of a virtual $W$ boson. The dominant Feynman diagram for each mode is shown in
Fig.~\ref{fig:feynmantbtqb}.

The single top production cross section has been calculated at next-to-leading order in QCD (NLO).
The NLO cross section for the s-channel process is $0.88^{+0.07}_{-0.06}$~pb~\cite{Smith:1996ij} at $\sqrt{s}
= 1.96$~TeV with $m_t = 175$~GeV. For the same settings, the NLO cross section for the t-channel process is
$1.98^{+0.23}_{-0.18}$~pb~\cite{Stelzer:1997ns,Harris:2002md}.

Searches for single top production were carried out in Run~I, but the
integrated luminosity of about 90 pb$^{-1}$ was not enough for an
observation. Upper limits on the cross sections were published by both
the {\dzero}~\cite{Abbott:2000pa,Abazov:2001ns} and
CDF~\cite{Acosta:2001un,Acosta:2004} collaborations. At the 95\%
confidence level, {\dzero}'s upper limit on the $s$-channel production
cross section is 17~pb and the CDF limit is 18~pb. At the same
confidence level, the upper limit on the $t$-channel production cross section is 22~pb by
{\dzero} and 13~pb by CDF. CDF also published a combined limit of 14~pb.

The final state for s-channel single top production consists of the decay products from the $W$ 
from the top quark decay, a b-quark jet from the top decay, and a b-quark jet produced with the top quark.
The final state for t-channel single top production consists of the decay products from the $W$ 
from the top quark decay, a b-quark jet from the top decay, and a light quark jet produced with the top.
Higher-order corrections can result in additional jets in both the s-channel and the t-channel, in particular
in an additional b-quark jet in the t-channel final state from the splitting of an initial state gluon into
a $b\bar{b}$ pair.
The experimental searches for single top production focus on the decay of the $W$ to an electron or a muon
since the all-hadronic channel has overwhelming backgrounds from QCD multi-jet events. 

This document describes a search for electroweak single top quark production using a data sample from Run~II 
at the Tevatron with the upgraded \dzero and CDF detectors. The data sample corresponds to an 
integrated luminosity of between 156pb$^{-1}$ and 169pb$^{-1}$ depending on the analysis channel and experiment.

\section{\dzero Search for Single Top Quark Production}
The search for single top quark production at \dzero starts with very loose event selection
designed to select events containing a $W$ and at least two jets out of the large background 
from QCD multi-jet production while keeping a high acceptance for single top events. The analysis 
is done separately for the electron and muon channel.
The event is required to contain exactly one isolated electron (muon) with $p_T>15GeV$ and $|\eta_{det}|<1.1$
( $|\eta|<2.0$), $\met>15GeV$, and between 2 and 4 jets with $p_T>15GeV$ and $|\eta_{det}|<3.4$. 
The leading jet must have $p_T>25GeV$ and $|\eta_{det}|<2.5$.
At least one of the jets is required to be identified as a b-quark jet using several b-tagging algorithms. 
Events with a soft muon tag (SLT) are analyzed separately from those with a lifetime tag to be able
to combine the results. Two lifetime tagging algorithms are applied, a) based on the reconstruction of a 
secondary vertex (SVT) and b) a jet lifetime probability algorithm (JLIP).
The two lifetime tag results are not combined because they are not based on independent datasets. Instead, 
the JLIP result is used as a cross-check of the combined SLT and SVT result.

The dominant background in the searches is from $W$+jets and $Z$+jets events, 
in particular $Wb\bar{b}$ production. 
Additional backgrounds are from misreconstruction of QCD multi-jet events and top pair production, 
both from the lepton+jets final state and the dilepton final state where one of the leptons was not reconstructed. 
There are also smaller contributions from di-boson events. The background due to misreconstructed multi-jet events 
is evaluated using a multi-jet data sample. The background distributions from to $W$+jets and $Z$+jets events as
well as di-boson events are modeled using data events passing preselection cuts but failing the b-tagging 
requirement. These are then normalized by the average jet tagging probability which is determined in a QCD 
multi-jet sample. 
The $t\bar{t}$ background is estimated from Monte Carlo, as is the single top signal acceptance.

\begin{figure}
\begin{minipage}{0.5\textwidth}
  \centering
  \psfig{figure=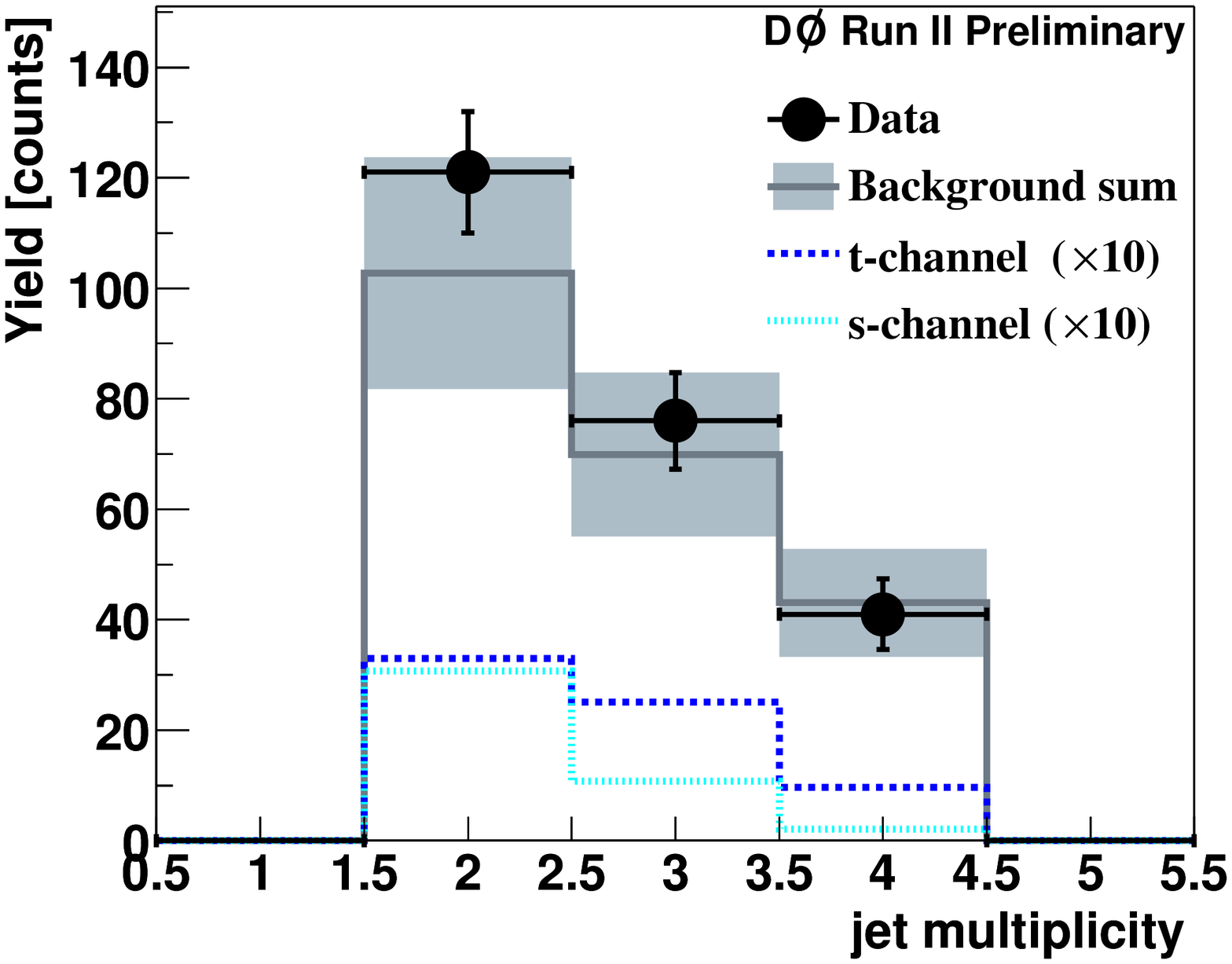,height=2.4in}
\end{minipage}
\begin{minipage}{0.5\textwidth}
  \centering
  \psfig{figure=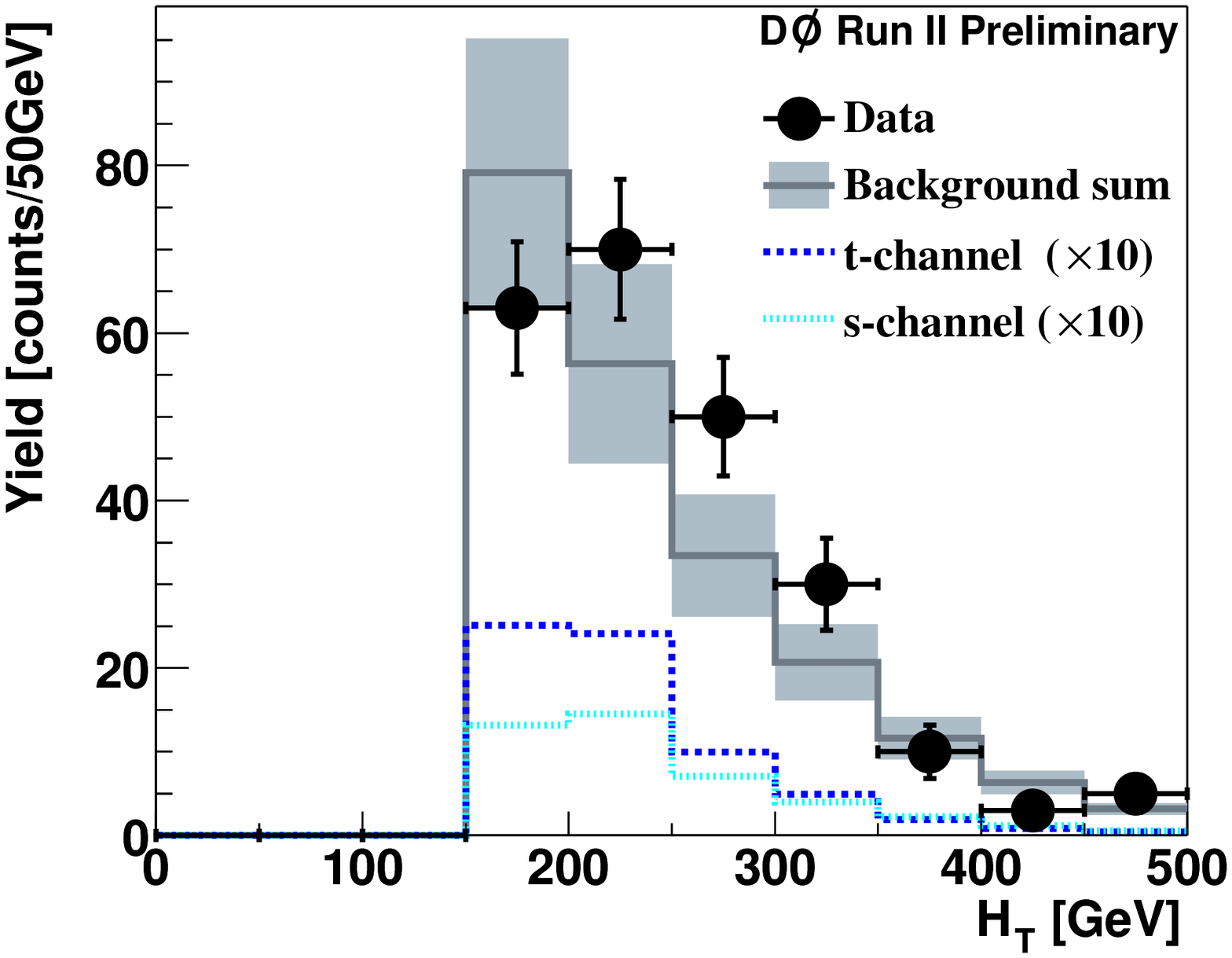,height=2.4in}
\end{minipage}
\caption{The jet multiplicity distribution (left) and event transverse energy distribution ($H_T$, right)
after the final selection cut for the \dzero data, combining the electron and muon channel and the SLT 
and SVT tagging results.
\label{fig:d0finaldistr}}
\end{figure}

A final selection cut is used to separate the single top signal events from the backgrounds. It was found that the
scalar sum of the transverse energies of the lepton, \met, and the leading two jets ($H_T$) 
gives good separation between the single top signal and the dominant background from $W$+jets. 
Fig.~\ref{fig:d0finaldistr} shows the jet multiplicity and the $H_T$ distribution after the final selection cuts,
comparing the data for the combination of SLT and SVT taggers to the sum of the background contributions.

The observed number of events after the final selection cut is consistent with the sum of the backgrounds 
within uncertainties. A 95\% C.L. limit is derived for the single top production cross section using Bayesian 
statistics and including systematic uncertainties. The largest contribution to the systematic uncertainties
are due to the jet energy scale, the modeling of b-tagging and the trigger, and the background
normalization. The limits (expected limits) are 19pb (16pb) 
for the s-channel, 25pb (23pb) for the t-channel, and 23pb (20pb) for the combination of s-channel
and t-channel.

\section{CDF Search for Single Top Quark Production}
The search for single top production at CDF has focused on the t-channel and on the combination of s-channel 
and t-channel. The basic selection cuts for both analyses are one isolated electron or muon with $p_T>20GeV$
and $|\eta|<1.0$, $\met>15GeV$, and exactly two jets with $p_T>15GeV$ and $|\eta|<2.8$. At least one of the
jets is required to be b-quark tagged with a secondary-vertex tagging algorithm.
The single top signal events are then selected by requiring that the invariant mass of the lepton, the
reconstructed neutrino, and the b-tagged jet ($M_{l \nu b}$) fulfills $140GeV<M_{l\nu b}<210GeV$. The t-channel
analysis makes the additional requirements that the leading jet has $p_T>30GeV$ and that
exactly one jet is b-tagged.


In order to increase the statistical sensitivity to the single top signal, a maximum likelihood fit to a
discriminant variable is performed. For the t-channel search, this variable is chosen to be the pseudo-rapidity
of the non-b-tagged jet (the light quark produced with the top quark), corrected for the charge of the lepton from 
the top decay. This distribution is peaked in the forward direction for t-channel signal events and is
symmetric around zero for all backgrounds.
For the combined s-channel and t-channel search, the fit variable is chosen to be the transverse event energy 
$H_T$ which has similar shape for the two signal contributions.

\begin{figure}
\begin{minipage}{0.5\textwidth}
  \centering
  \psfig{figure=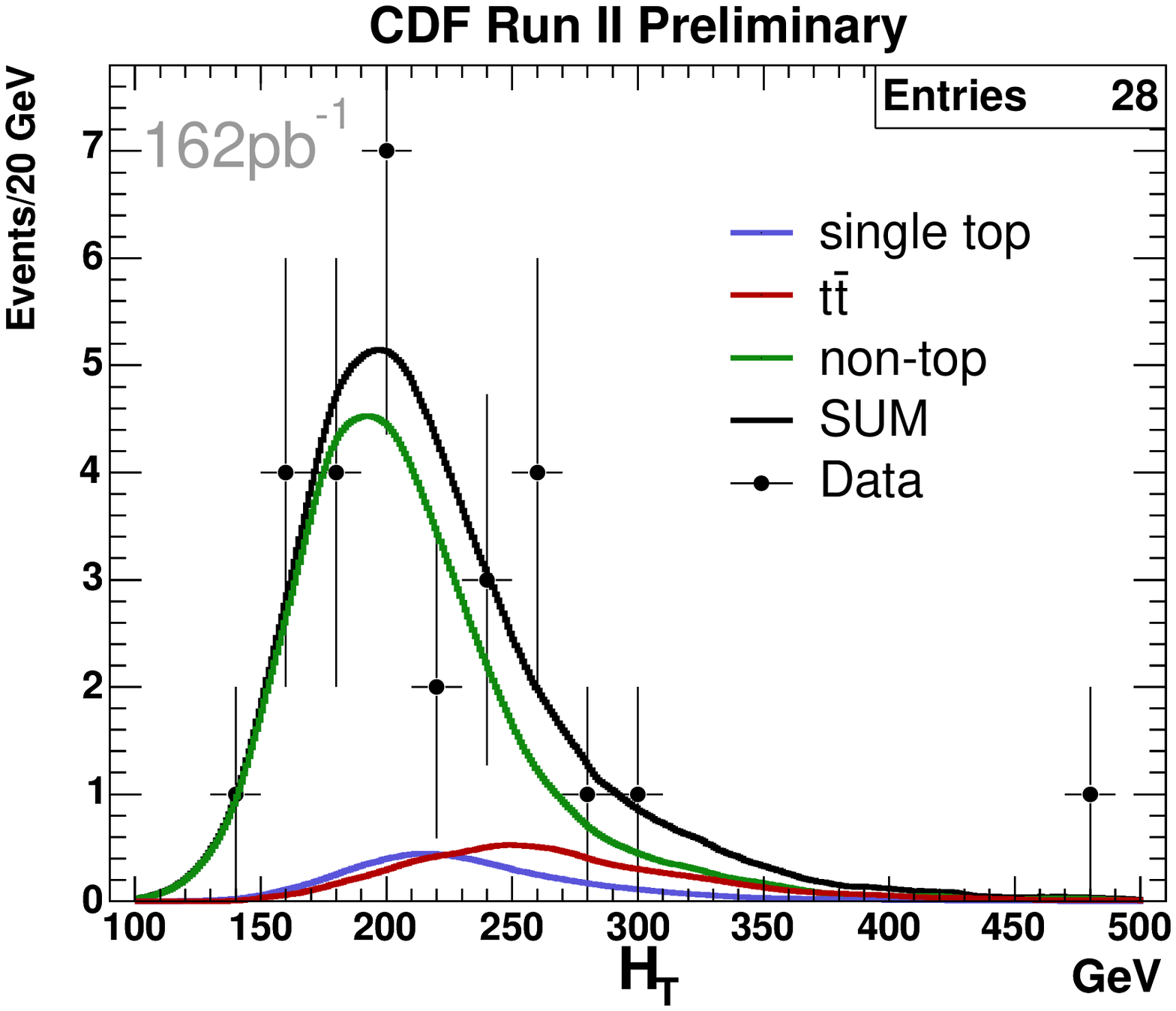,height=2.5in}
\end{minipage}
\begin{minipage}{0.5\textwidth}
  \centering
  \psfig{figure=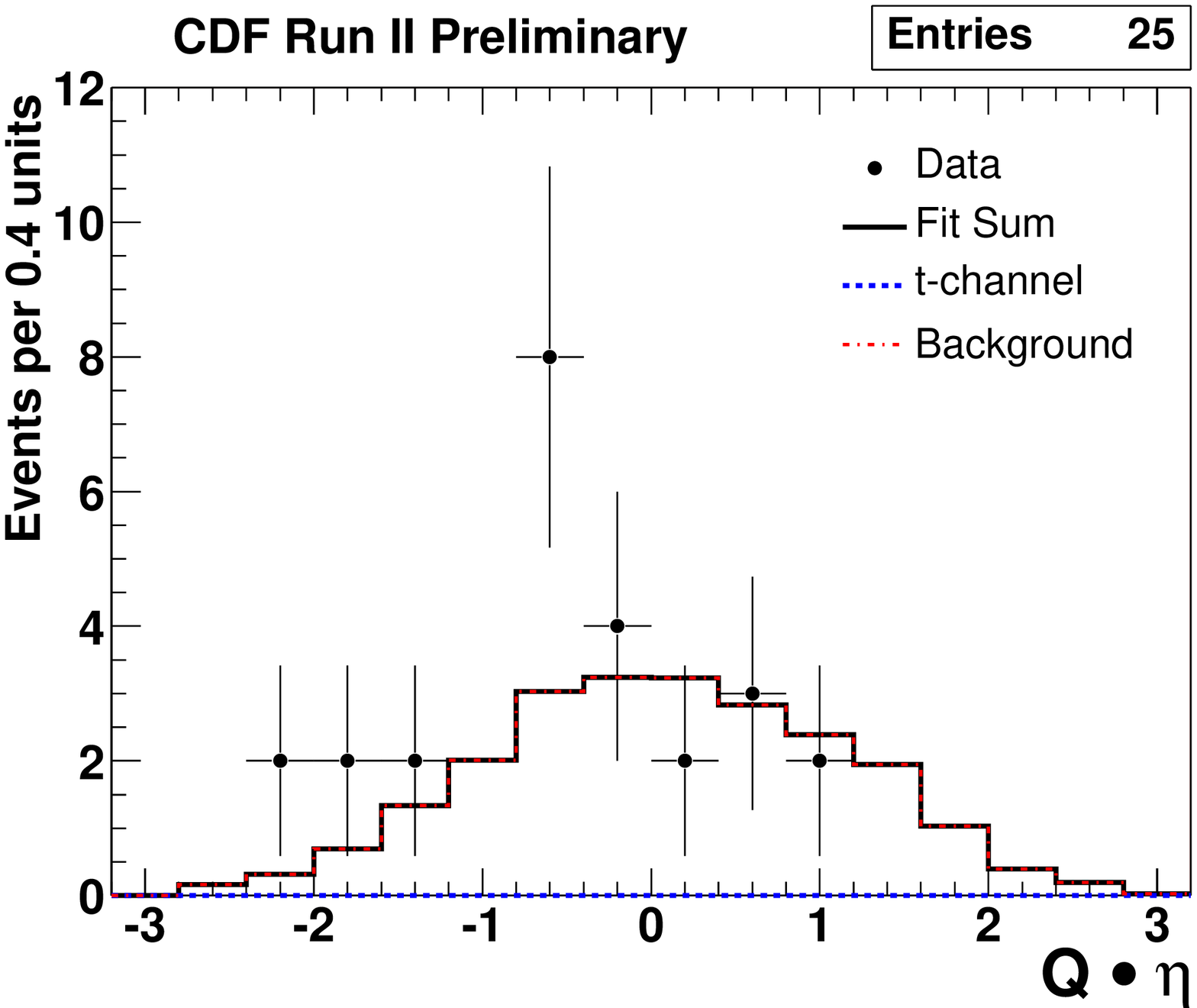,height=2.5in}
\end{minipage}
\caption{The result of the template fit to the CDF data, for the combined s-channel and t-channel search
(left) and the separate t-channel search (right).
\label{fig:cdffinaldistr}}
\end{figure}

The result of the template fit to the CDF data is shown in Fig.~\ref{fig:cdffinaldistr}. Both distributions
are consistent with the sum of the backgrounds and show no evidence for a single top signal. The resulting
95\%C.L. limits (expected limits) including systematic uncertainties are 13.7pb (14.1pb) for the 
combination of s-channel and t-channel and 8.5pb (11.3pb) for the t-channel.

\section{Summary}
We have shown preliminary results for the Tevatron Run~II searches for electroweak production of 
single top quarks for a dataset of approximately 160pb$^{-1}$. Both experiments find good agreement between 
the expected backgrounds and the observed data and both set 95\% C.L. limits on the 
single top production cross section. Observation of single top events is expected in the future
with increased luminosity and improved analysis methods.


\end{document}